\documentclass[a4paper]{jpconf}
\usepackage{graphicx}
\usepackage[frozencache]{minted}
\begin{document}
\title{Using a DSL to read ROOT TTrees faster in Uproot}

\author{Aryan Roy$^1$ and Jim Pivarsk$i^2$}

\address{$^1$ Manipal Institute of Technology, Manipal, India}
\address{$^2$ Princeton University, Princeton NJ, USA}
\ead{aryanroy5678@gmail.com, pivarski@princeton.edu}
\begin{abstract}
Uproot reads ROOT TTrees using pure Python. For numerical and (singly) jagged arrays, this is fast because a whole block of data can be interpreted as an array without modifying the data. For other cases, such as arrays of \mintinline{c++}{std::vector<std::vector<float>>}, numerical data are interleaved with structure, and the only way to deserialize them is with a sequential algorithm. When written in Python, such algorithms are very slow.

We solve this problem by writing the same logic in a language that can be executed quickly. AwkwardForth is a Domain Specific Language (DSL), based on Standard Forth with I/O extensions for making Awkward Arrays, and it can be interpreted as a fast virtual machine without requiring LLVM as a dependency. We generate code as late as possible to take advantage of optimization opportunities. All ROOT types previously implemented with Python have been converted to AwkwardForth. Double and triple-jagged arrays, for example, are 400× faster in AwkwardForth than in Python, with multithreaded scaling up to 1 second/GB because AwkwardForth releases the Python GIL. We also investigate the possibility of JIT-compiling the generated AwkwardForth code using LLVM to increase the performance gains. In this paper, we describe design aspects, performance studies, and future directions in accelerating Uproot with AwkwardForth.
\end{abstract}

\section{Introduction} 
A majority of the particle physics data today is stored as ROOT files\cite{BRUN199781}. However, there is considerable variation in how the ROOT files are serialized. The newer versions (such as RNTuple) are exclusively stored in a columnar format. This means that that the data for each column is stored as a contiguous block. This is not entirely true for the older format, TTree, which stores simple data types in a columnar fashion and complex data types in a record-oriented format, which means that each row is stored as a contiguous block with bytes between rows. This introduces a need to iterate over objects when deserialising the data.

Only a handful of libraries in Python allow users to read ROOT files, and only one allows them to do so without any compiled code dependency, that library is Uproot \cite{Pivarski_Uproot_2017}. Uproot is a part of the Scikit-HEP ecosystem with a ROOT file reader implemented in pure Python. When it comes to TTrees, Uproot is much slower when deserialising record-oriented data, such as data types of nested depth greater than one. This is due to the fact that data with nested depth less than or equal to one (numerical and singly jagged) can be read by casting whole blocks of data as NumPy arrays, making it a constant time operation. For data types of nested depth greater than one (doubly jagged and more), the Pythonic deserialiser has to alternate between reading list contents and list lengths. This slows down the Pythonic deserialiser by orders of magnitude.
In this paper, we present a solution to this problem by updating the Uproot library to generate specialised deserialisation code in a Domain Specific Language (DSL)  to read each unique data type. The new DSL based deserialiser is the default option starting from Uproot version 5.

\section{Design of the DSL}

The design of the DSL informs the majority of the subsequent design decisions of the new deserialiser. The biggest choice to be made is between a compiled language and an interpreted one. While a compiled language could deliver higher performance, it would also introduce the need for a runtime compiler, which would be hard to install as a dependency. On the other hand, an interpreted language, while easy to install and use, could suffer from lower speed when compared to the compiled alternative.

Another important restriction governing the choice of DSL is our inability to simply pre-compile the deserialisation code for the specific data types. This is because the data types in the ROOT file format are discovered at runtime by reading the TStreamerInfo part of the file. This means that the code for each type of data will necessarily  have to be generated at runtime.

Given the fact that even developers do not see the code generated in the DSL, it does not need to be highly human-readable. This opens us up to the possibility of choosing a DSL that is instead easy to generate.

To summarise, the restrictions for the choice of DSL are as follows:
\begin{itemize}
    \item It should be lightweight, i.e.\ should not depend on compilation tool chains like LLVM \cite{LLVM:CGO04}.
    \item It should be considerably faster than Python.
    \item It does not need to be easy to read but it does need to be easy to generate. For example, the syntactic indentation of Python is not necessary and hard to generate consistently.
\end{itemize}
\section{AwkwardForth DSL}
AwkwardForth \cite{pivarski2021awkwardforth} is a DSL designed to satisfy the requirements listed in the last section. It is based on standard Forth with some additional built-in commands for parsing files. It is shipped with Awkward Array \cite{Pivarski_Awkward_Array_2018}, a Python library for handling nested lists of arbitrary lengths, already required by Uproot to represent the complex data types serialised in ROOT files, since NumPy arrays are limited to purely numeric data.

AwkwardForth is considerably faster than Python. It takes about 5--10~ns per instruction to execute compared to about 1000--2000~ns per instruction for Python. This higher speed can be primarily attributed to two factors:
\begin{itemize}
    \item Python folows object pointers at runtime, AwkwardForth, like all Forths, only has one data structure, a stack of integers.
    \item Python checks types at runtime, AwkwardForth has only one type, either all 32-bit integers or all 64-bit integers.
\end{itemize}
Forth has a minimal syntax, consisting only of a stream tokens seperated by whitespace. All of these properties are retained by AwkwardForth, making it well suited for our task.

The Awkward Array library ships with a Virtual Machine (VM) for running AwkwardForth code. This VM requires no extra installation and is pip installed along with the rest of the library, so there is no extra dependency requirement.

While not having a compiler makes the new reader more accessible, we are aware of the fact that some of the potential users could already have a JIT-compiler like Numba \cite{lam2015numba} installed. In principle, we could use Numba if it is available in the environment to JIT-compile the AwkwardForth code. This would give an option for those users to have an even faster reader than the interpreted one. This possibility is discussed in a later section.

\section{Data Types Covered}
Many primitive and complex data types have been encountered by Uproot users in its 6-year history. The new deserialiser must be capable of reading all of these data types to be the default option for Uproot version 5. To realise this, we implemented support for a large number of frequently used data types, including all of the STL types (vectors, maps, sets etc), arrays, a number of built-in classes (TString, Tarray, TDatime, TRefArray etc.), and user-defined classes.

The only data types not implemented in AwkwardForth are those that can't be represented as Awkward Arrays and class features that have never been reported in Uproot's history. These cases fall back on the old Pythonic reader.

\section{The Implementation}
Implementing the reader required making a number of design choices to ensure that it covers all the corner cases in a complex file format like ROOT.

Reading a ROOT TTree file requires discovering a lot of information at runtime. The same Awkward data type may be serialized as different C++ classes, for example, strings can be represented as std::string, char* and TString. This variety within the same data type is due to historical reasons and the richness of the C++ type system. 

But sometimes, even the same C++ data type can be serialized in different ways. For example, sometimes object headers are skipped, and the decision to skip some headers are encoded in other headers in the data stream, not in external metadata. If all code is generated before deserializing, decisions about which headers to skip would have to be made repeatedly for all objects in the data stream. While that would have some impact on fully JIT-compiled code (preventing vectorization), it has more impact on interpreted code like AwkwardForth.

To be most effective, we need to generate the AwkwardForth code as late as possible, after we know which object headers exist. Object headers are represented by an AwkwardForth "skip" command and missing headers involve no code at all. Neither case involves expensive code branching ("if" statements).

We implement this late AwkwardForth generation as a new feature of the old Pythonic deserialiser. The Pythonic deserialiser runs over the first data entry, generating code as a side-effect. Then the AwkwardForth runs over the whole dataset (repeating the first entry).

Due to the GIL in Python, the old Pythonic deserialiser could not make effective use of multi-threading. This was a huge disadvantage as multi-threading can scale up the reading of large files. The new AwkwardForth based deserialiser is not limited by the GIL as the VM is written in C++. This allows multi-threaded execution of the AwkwardForth code, enabling the deserialisation of large files at a much higher rate. While the execution of the AwkwardForth code can be multi-threaded, the code generation itself is done in a single thread, to avoid redundantly generating the same code in multiple threads. When code-generation is done, the completely formed VM is distributed to all the threads, then they can each start reading the file from different points in the bytestream.




\section{Performance}

\begin{figure}
\centering
\includegraphics[width=0.5\linewidth]{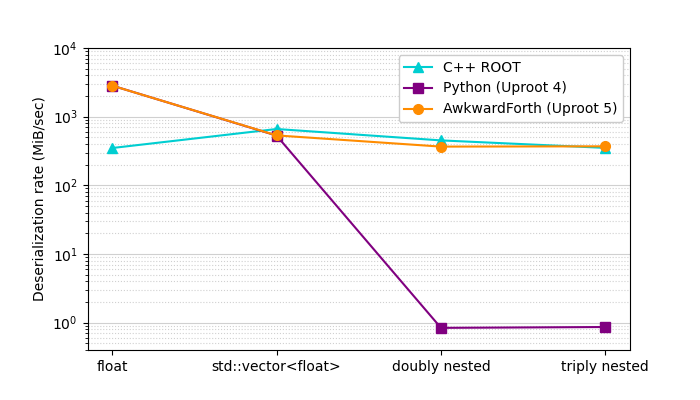}\hspace{2pc}%
\caption{\label{fig:1}Deserialization rate of the new deserializer, the old Pythonic deserializer, and C++ ROOT (C++ \mintinline{c++}{for} loop calling \mintinline{c++}{TTree::GetEntry}).}
\end{figure}

The new deserialiser performed as well as expected. Figure~\ref{fig:1} shows the deserialisation rate of large samples of these four data types: \mintinline{c++}{float}, \mintinline{c++}{std::vector<float>}, \mintinline{c++}{std::vector<std::vector<float>>}, that is, doubly nested list, as well as triply nested lists.

The first two of these are columnar, and therefore unaffected by the update. The last two, however, are examples of record-oriented data, which the new AwkwardForth deserialiser reads approximately 400$\times$ faster than the old Pythonic deserialiser.

\begin{figure}
\centering
\includegraphics[width=0.45\linewidth]{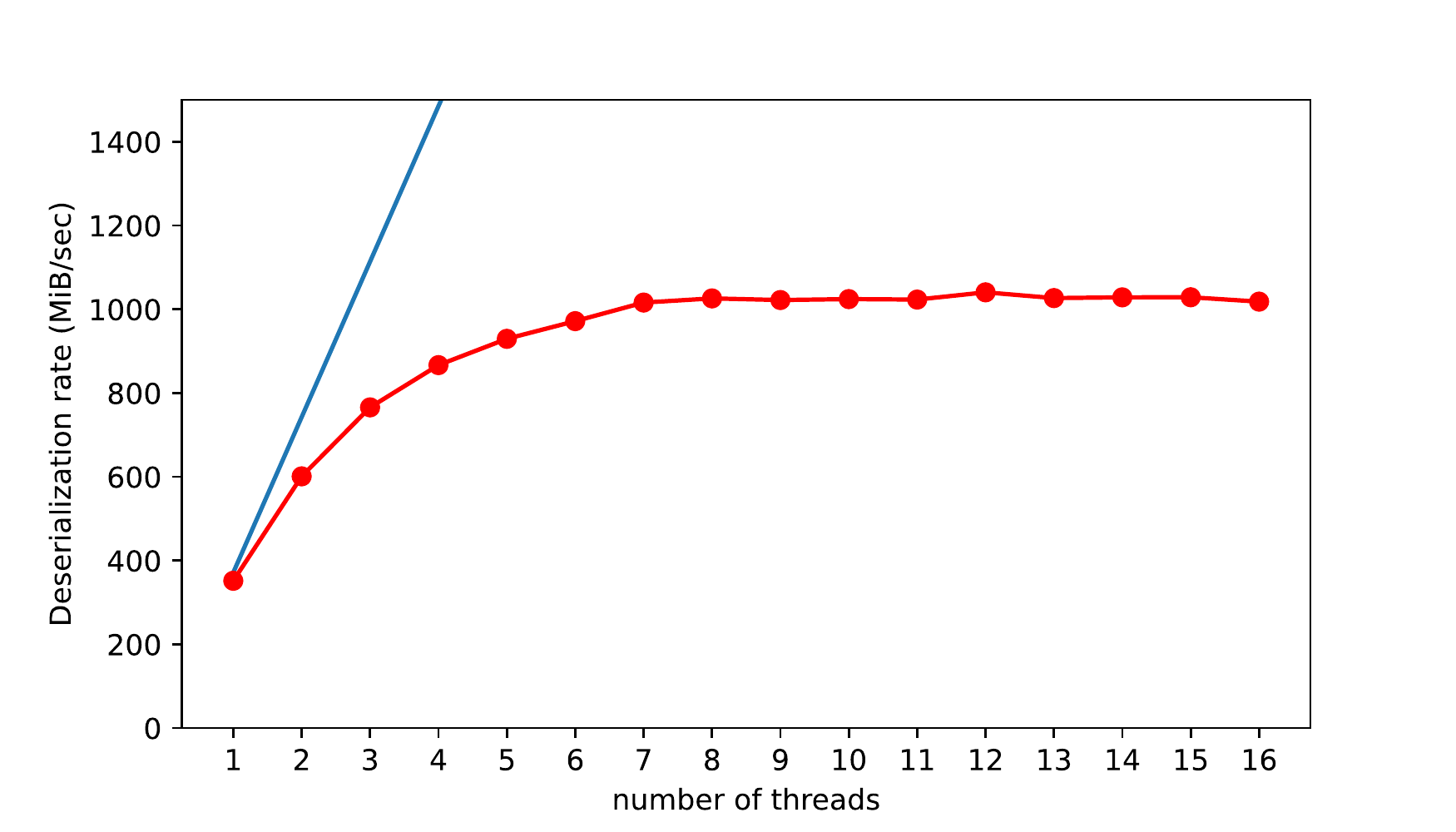}\hspace{2pc}\includegraphics[width=0.45\linewidth]{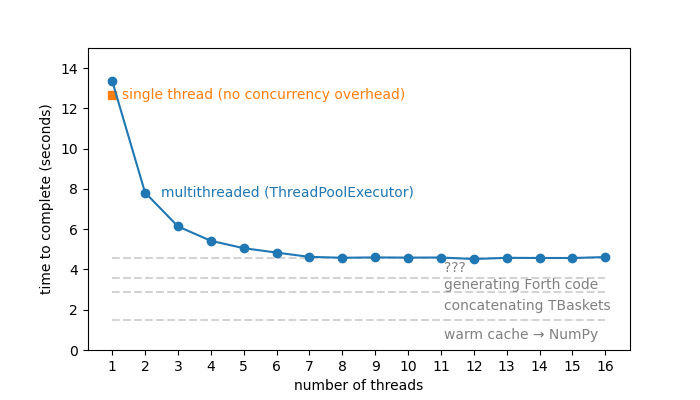}\hspace{2pc}%
\caption{\label{fig2}Scaling of the new deserialiser with the number of threads. Left: vertical axis is rate. Right: vertical axis is time.}
\end{figure}

In Figure~\ref{fig2}, we can see that the deserialisation rate also scales very well with the number of threads (up to about 1000~MiB/sec). After which the serial part dominates the end to end workflow, in accordance with Amdahl's law. Before and after the parallel part of the workflow, data need to be copied into the Python process, the AwkwardForth code needs to be generated, finalized arrays needs to be concatenated, etc. We could identify about 80\% of the flat part of the curve as serial steps.

All of these tests were carried out with uncompressed data that was already loaded from disk into the operating system's virtual memory (RAM). Thus, the rates measured do not include decompression (typically slower) or disk-reading (depends on the speed of the disk).


\section{Future Directions: JIT-compiling AwkwardForth}
The VM used to run the AwkawardForth code consists of a minimal interpreter written in C++, which interprets virtual bytecode in a loop. This was sufficient for this task as most of the AwkwardForth code generated to deserialise ROOT TTrees consists of simple stack manipulations, type conversions, and endian swapping. The lightweight VM without any code optimisation features is able to execute the AwkwardForth code very efficiently and generate favourable results as mentioned in the last section.


\subsection{JIT-compiling AwkwardForth}
One of AwkwardForth's design goals was to not need a compilation toolchain like LLVM, but since Numba is a popular JIT-compiler in the scientific Python world, it may be available anyway. In principle, we could check to see if Numba is installed, and if it is, fully JIT-compile the AwkwardForth code. This way, the same AwkwardForth code could be used with and without JIT-compilation; JIT-compilation would only provide an optional speed boost.

In this section, we investigate the extent of performance gains that could be achieved in such a case. As a proof of concept, we implemented a bare-bones compiler, written in Python, to compile AwkwardForth bytecode to LLVM. This new experimental compiler was designed to fit into the existing deserialisation pipeline of Uproot. We reuse the tokenizer and the bytecode gnerator of the existing interpreter and translate this bytecode into Python code that Numba can compile (which we'll be calling ``Numba code''). This allows us to use code that is already tested and deployed as a first step in the compilation workflow.

The new compiler works by stepping through the list of bytecodes provided by the VM and generating the equivalent Numba code. The Numba code is then put into a function with a Numba \mintinline{python}{@numba.jit} decorator to compile it. The functions for pushing and popping from the stack are predefined Numba functions available in the Python environment. The stack is implemented as a NumPy array with a fixed length and an integer cursor to keep track of the head of the stack. These are both passed into the generated function as arguments. This allows us to manipulate the stack from outside, just as we can with the current AwkwardForth implementation. 
The task of deserialising typically requires execution of simple commands (with minimal branching) inside a loop. This same set of commands are repeated for each entry in the input dataset, which can be arbitrarily large. Since LLVM optimizes the code it is given, we can write naive code. However, it does help if the Numba code has fewer checks for stack underflow and overflow, and we can usually identify such situations while generating the Numba code by counting pushes and pops. In some cases, it's not possible to know the number of elements that could be be pushed or popped, and runtime error checks are left in for only these cases.

\subsection{The Performance of the new compiler}
We implemented a few operations from the VM and ran identical pieces of AwkwardForth code in both the interpreted VM and the new experimental compiler. The code that we tested does not include any data parsing, so that we can measure the rate of computation independently of fetching data from RAM. Thus, these tests represent the best possible speedup from JIT-compilation, for code that is arithmetically intense.

Our test computation is

\vspace{-0.75 cm}\begin{eqnarray}
 & x_{i + 1} = (x_i + 1) * (x_i - 2) + 3 \\
 & \mbox{AwkwardForth: \mintinline{Forth}{dup 1 + swap 2 - * 3 +}}
\end{eqnarray}

\noindent Including the \mintinline{forth}{do}\ldots \mintinline{forth}{loop} itself, this is 10~AwkwardForth instructions, which we expect to be 5--10~ns per instruction. When JIT-compiled (at \mintinline{bash}{-O3}) in x86, the expression becomes

\vspace{-0.1 cm}\begin{equation}
\begin{tabular}{l l}
\tt lea & \tt eax, [rdi+1] \\
\tt sub & \tt edi, 2 \\
\tt imul & \tt eax, edi \\
\tt add & \tt eax, 3 \\
\end{tabular}
\end{equation}

\noindent The expression only uses the fastest hardware commands (addition, subtraction, and multiplication, each about 1~clock tick) and in a way that they cannot be optimized away.

\begin{figure}
\centering
\includegraphics[width=0.47\linewidth]{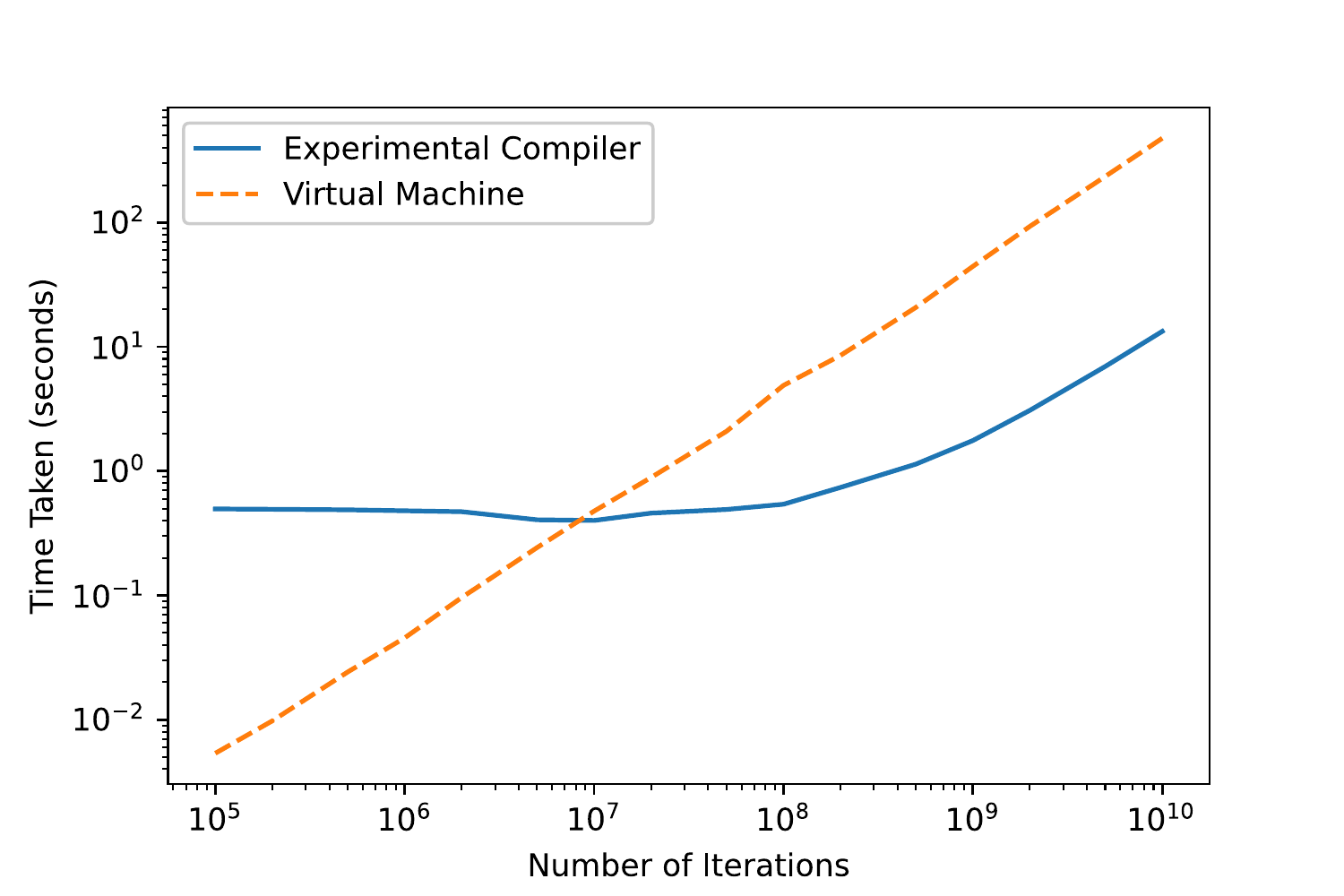}\hspace{2pc}%
\caption{\label{fig3}Time spent in computation, using the (proof of principle) JIT-compiler and the interpreted VM.}
\end{figure}

Figure~\ref{fig3} shows the results of this comparative study. JIT-compilation adds a constant-sized overhead of about 400~ms for the compilation itself, such that it only begins to scale with $\sim$10$^7$ iterations or more. However, in the asymptotic limit of many iterations, the JIT-compiled version has a $\sim$30$\times$ faster rate.

The above test excluded the cost of data transfer, which is highly variable. Different computers can vary from 1~GB/sec to 10~GB/sec, and the number of bytes per iteration depends on the type of data structure and data values (lists with many items versus few items). Record-oriented data has at least tens of bytes per entry, so the bottleneck due to data transfer would be at least 1~second for 10$^9$ entries (assuming 10~bytes per entry and 10~GB/sec), which is about the same rate as the experimental compiler itself. If data transfer is slower than this (more bytes per entry or slower memory transfer rate), then data transfer would dominate over the gains from JIT-compilation. Thus, the benefit of JIT-compilation would depend strongly on use-case.

From this study, we have learned that JIT-compilation of the AwkwardForth would be at least sometimes beneficial, by as much as 30$\times$ in the best case. However, data transfer rates can mask this difference in some cases and not others.

\section{Conclusion}

In this paper, we presented the implementation details of a DSL based ROOT file deserialiser. The new deserialiser uses the AwkwardForth DSL to generate type-specific code to read ROOT TTrees files. The new deserialiser was implemented to overcome the orders-of-magnitude slowdown of using Python to deserialise data in sequential loops for record-oriented data. The new deserialiser showed a 400$\times$ gain in speed for this type of data.

We also studied the possibility of optionally JIT-compiling AwkwardForth code when Numba is in the Python environment. A proof of principle implementation showed a 30$\times$ speedup, in an ideal case of arithmetically intensive instructions. Data transfer rates, however, can mask this difference when more than tens of bytes per entry need to be transferred.

Starting in Uproot version 5 (already released), AwkwardForth became the default deserialiser. All up-to-date users of the Uproot package with complex data to read are now enjoying its benefits.

\section{Acknowledgements}

This work was supported by the National Science Foundation under Cooperative Agreement OAC-1836650 (IRIS-HEP).

\section{References}
\bibliographystyle{vancouver}
\bibliography{mybib}

\end{document}